\documentclass[12pt]{article}
\usepackage{amssymb}

\usepackage{amsmath}

\begin{document}

\begin{center}
\vspace{1cm}{\Large {\bf On an (Interacting) Field Theories With
Tensorial Momentum}}

\vspace{1cm} {\bf Ruben Mkrtchyan} \footnote{ E-mail: mrl@r.am}
\vspace{1cm}

\vspace{1cm}

{\it Theoretical Physics Department,} {\it Yerevan Physics
Institute}

{\it Alikhanian Br. St.2, Yerevan, 375036 Armenia }
\end{center}

\vspace{1cm}
\begin{abstract}
The construction of field theories with space-time symmetries,
including tensorial charges (i.e. of M-theory type), initiated in
hep-th/9907011, is extended to include interaction. For SO(2,2)
gravity in a tensorial space-time, with space-time symmetry
consisting of Lorentz generators and "translations", represented
by second-rank antisymmetric tensor, the cubic interaction terms
are constructed by requirement of maintaining the gauge invariance
property of theory. This interaction is essentially unique.
\end{abstract}

\renewcommand{\thefootnote}{\arabic{footnote}} \setcounter{footnote}0
{\smallskip \pagebreak }

\section{Introduction}

Space-time algebras with tensorial momentum appeared in study of
branes theory, see \cite{Tow} for a review. The most general such
algebra is that of M-theory, where anticommutator of supercharges
includes all possible tensors:

\begin{eqnarray}
\left\{ \bar{Q},Q\right\} &=&\Gamma ^{i}P_{i}+\Gamma
^{ij}Z_{ij}+\Gamma ^{ijklm}Z_{ijklm},  \label{eq1}\\
  i,j,...
&=&0,1,2,..10. \nonumber
\end{eqnarray}

The natural approach from the point of view of modern field theory
is to try to construct the field theories, invariant w.r.t such
(super)-algebras. Such an approach was initiated in \cite{Man1}.
There we didn't address the problem in whole generality, but make
the following simplifications. First, as noted in \cite{Bars},
algebra (\ref{eq1}) can be rewritten in a manifestly SO(2,10)
invariant form:

\begin{eqnarray}
\left\{ \bar{Q},Q\right\} &=& \Gamma ^{\mu \nu }P_{\mu \nu
}+\Gamma
 ^{\mu \nu \lambda \rho \sigma \delta }Z_{\mu \nu
\lambda \rho \sigma \delta }^{+},
\label{eq11} \\
\mu \nu ,... &=&0^{\prime },0,1,...10 \nonumber
\end{eqnarray}

As seen, energy-momentum vector disappear from the r.h.s, namely
it joins with 11D second rank tensor into 12D second rank tensor.
Then we took the bosonic part of (\ref{eq11}) (with Lorentz
generators), with non-zero second-rank tensor only, and considered
that algebras at different lower dimensions. So, the algebra we
shall consider consists of the following set of generators:

\begin{eqnarray}
M_{\mu \nu}, P_{\mu \nu} \label{eq111} \\ \mu, \nu=0^{\prime },
0,1,...q \nonumber
\end{eqnarray}
where $M_{\mu \nu}$ are SO(2,q) Lorentz generators, $P_{\mu \nu}$
are Abelian "translations". The first step in construction of
(interacting) field theories is construction of the unitary
irreducible representations of algebra (\ref{eq1}) on the language
of relativistic fields equations. Such an equations are
constructed in \cite{Man1}, \cite{Man2}. The usual principles of
construction of such a representations are applicable for more
general cases than usual Poincare algebra, particularly in a given
case. The simplest representation is in the space of scalar
function on space, dual to momenta in (\ref{eq1}), selected by
Klein-Gordon type equations. Back in momentum representation they
are:

\begin{eqnarray}
(TrP^{2}-2m_{1}^{2})\Phi (P_{\mu \nu }) &=&0,  \label{eq2} \\
(TrP^{4}-2m_{2}^{4})\Phi (P_{\mu \nu }) &=&0,   \label{eq22}\\
&...&  \nonumber
\end{eqnarray}
In a similar way the equations of  reduced spin 1/2 can be
constructed. The novelty appears for spin 1 equation, which can be
expected to incorporate some generalization of notion of gauge
invariance. It appears that statement of gauge symmetry for the
theories considered have to be generalized into the form

\begin{equation}
\delta _{1}S_{1}+\delta _{2}S_{2}+....+\delta _{n}S_{n}=0
\label{eq3}
\end{equation}
where  $S_{1}$, $S_{2}$ ... are actions for equations,
corresponding to Eqs. (\ref{eq2}), (\ref{eq22}), ... for spin 1,
and  $\delta _{1}$, $\delta _{2}$, ... are corresponding
variations. In this report we shall consider the spin 2 case in
dimension 2+2. The exact equations for that case are the
following. The first level action $S_{1}$ is
\begin{equation}
\begin{array}{l}
S_1 = \int dx( - \partial _{\nu \kappa } h_{\alpha \beta }
\partial _{\nu \kappa } h_{\alpha \beta }  + 4\partial _{\beta
\kappa } h_{\alpha \beta } \partial _{\nu \kappa } h_{\alpha \nu}
- \partial _{\beta \kappa } h_{\alpha \beta } \partial _{\nu
\alpha } h_{\kappa \nu }  \\
 + 4\partial _{\alpha \beta }
h_{\alpha \kappa } \partial _{\beta \kappa } h +
\partial _{\beta \kappa } h \partial _{\beta
\kappa } h) \label{L1}
\end{array}
\end{equation}
with $h=h_{\mu}^{\mu}$.  This expression is unique among those of
second order over derivatives and over field $h_{\mu \nu }$, which
goes into the quadratic part of the General Relativity Lagrangian
after reduction, i.e on the mass shell of higher level equations
(see \cite{Man1}).

Second level action $S_{2}$ is
\begin{equation}
\begin{array}{l}
 S_2  = \int {dx( - 4\partial _{\alpha \beta } h_{\beta \kappa }\partial _{\kappa \lambda } \partial _{\lambda \mu } \partial_{\mu \nu } h_{\alpha \nu } }
 - 2\partial _{\alpha \beta } h_{\beta \mu } \partial _{\kappa \lambda } \partial _{\kappa \lambda } \partial _{\mu \nu } h_{\alpha \nu }  \label{L2} \\
- s(\frac{1}{2}\partial _{\kappa \lambda } \partial
_{\lambda\mu}h_{\alpha \beta } \partial _{\mu \nu }
\partial_{\nu\kappa } h_{\alpha \beta }  - \frac{1}{4}\partial _{\kappa \lambda } \partial _{\kappa \lambda } h_{\alpha \beta } \partial _{\mu \nu } \partial _{\mu \nu } h_{\alpha \beta } )) \\
 \end{array}
\end{equation}
where $s$ is an arbitrary real parameter, which can be fixed from
the requirement that 2+2 theory should be a part of 2+4 theory, as
shown in \cite{Man2}. This theory has the following gauge
invariance
\begin{eqnarray}
\delta _1 S_1  + \delta _2 S_2  = 0 \label{22}
\end{eqnarray}
with  variations, in momentum representation:
\begin{eqnarray}
\delta _{1}h_{\mu \nu } &=&(p^{2}\xi +\xi p^{2}-s p \xi p)_{\mu \nu } ,\label{d1}\\
\delta _{2}h_{\mu \nu } &=&\xi_{\mu \nu } \label{d2}
\end{eqnarray}
where gauge transformation parameter $\xi_{\mu \nu }$ is symmetric
tensor.

 Our main aim in this report is construction of first
non-trivial interaction term for the theory (\ref{L1}),
(\ref{L2}). The basis of construction will be the principle of
maintaining the (deformed) gauge transformation (\ref{22}). In
other words, we shall use Noether procedure, generalized to the
case of gauge invariance with few Lagrangians (\ref{22}).
    The construction of field theories in space-times with
coordinates, corresponding to (some) of tensorial charges, was
first, at best of our knowledge, addressed in \cite{Man1}. The
particle models in such spaces (with all tensor coordinates
activated) are constructed in a number of papers, see \cite{Rud},
\cite{Band}. In the beautiful paper \cite{Vas} a free field
equations are constructed  in the space of second-rank symmetric
tensor coordinates for OSp type algebras, which (equations)
describe the whole tower of higher spin fields, and analysis of
properties of these equations is given. Differences and
similarities with present approach are, first, that we are
constructing Lagrangians and actions, while in \cite{Vas} the
equations of motion are suggested, although, the number of
equations exceeds the number of field, which is counterpart of
necessity of few Lagrangians for one field in our approach.
Second, we are constructing a field theories for one irrep of
tensorial Poincare, while \cite{Vas} contains, as mentioned, the
whole tower of higher spins. See also references in \cite{Vas} for
an earlier ideas on field theories in a tensorial space-time.

\section{Cubic Interaction Terms For 2+2 Gravity}

So, we have to write down all possible "next order" over
$h_{\mu\nu}$ terms, which are first non-trivial, cubic,
interaction terms in both actions (\ref{L1}), (\ref{L2}) and in
transformations (\ref{d1}), (\ref{d2}), with arbitrary
coefficients, then calculate the variations and require the
fulfilment of gauge invariance equation (\ref{22}) in
corresponding order. This procedure will give a set of linear
equations on coefficients of different terms in next order actions
and variations. That system will be a strongly overdetermined, so
one can't guarantee the existence of solution. Indeed,
calculations for the simplest possible case of Chern-Simons
Lagrangian with linear over derivatives variations were
unsuccessful \cite{Mkr}. So, we find encouraging the fact, that in
this case the non-trivial solution exists. Moreover, as we shall
see below, in some sense that solution is unique.
    Now we shall describe the derivation of solution and present an
exact formulae.
    First, to the action $S_{1}$ the terms can be added of the
form $hhh\partial\partial$ where derivative are implied to be
acting on fields, and indexes should be contracted in some way.
There are many ways of constructing such terms, but integrating by
parts leads to the 19 independent terms. Some terms can be
excluded due to the finite dimensionality - 4, of range, run by
indexes, so that antisymmetrization over 5 of them is identically
zero. Direct check gives exactly one relation between mentioned 19
terms, so that we have precisely 18 independent 3-rd order terms
to be added to $S_{1}$. Next, corresponding terms should be added
to variation (\ref{d1}), namely of the type $\xi h
\partial\partial$, where, again, derivatives are implied to be
acting on a field $h $ or parameter $\xi$, and indexes are
contracted in appropriate way. There are 67 possible terms in the
first variation $\delta_{1}$, correspondingly another 67 unknown
parameters are introduced. Turning to the second Lagrangian, the
similar considerations lead to the 148 coefficients of the terms
of type $hhhh\partial\partial\partial\partial$, added to second
Lagrangian, and three terms of type $h\xi$ added to second
variation, i.e. the variation (\ref{d2}). There are relations
among mentioned 148 terms, similar to that in $S_{1}$, coming from
the finiteness of range run by indexes. Direct calculation with
the use of "Mathematica" shows existence of 48 such relations.
    So, we have to substitute variations into actions and check
the relation (\ref{22}) up to the order $\xi hh
\partial\partial\partial$. Direct calculation  gives the 9-parametric solution for the
overdetermined system of linear equations (\ref{22}). One of these
parameters corresponds to solution with zero variation of second
action, and variation of first one is proportional to its equation
of motion. Three parameters are appearing as redefinition of
parameter $\xi \rightarrow \xi h$ in free actions'
gauge-invariance statement (\ref{22}) with (\ref{L1}), (\ref{L2}),
(\ref{d1}), (\ref{d2}). Two parameters are also appearing from
free actions gauge-invariance statement through redefinition of
field $h\rightarrow hh$. All these solutions are in a reasonable
sense trivial. Among remaining three parameters two have a
property that in corresponding solutions all coefficients of third
order terms in $S_{1}$ are zero, although in $S_{2}$ third-order
terms are truly non-zero, so these solutions are "half
non-trivial". Finally, one parameter requires fully non-trivial
third-order interaction terms in both actions. That solution is
presented below. We present only next order terms w.r.t. the
(\ref{L1}), (\ref{L2}), (\ref{d1}), (\ref{d2}).

\begin{equation}
\begin{array}{l}
 S_1^{(3)}  = g\int {(2h_{\kappa \nu } \partial _{\nu \pi } h_{\xi \kappa } \partial _{\pi \alpha } h_{\alpha \xi } }
  + 2h_{\alpha \xi } \partial _{\nu \pi } h_{\xi \kappa } \partial
_{\pi \alpha } h_{\kappa \nu }   \\+ h_{\kappa \nu } \partial
_{\xi \kappa } h_{\alpha \xi } \partial _{\pi \alpha } h_{\nu \pi
}  +
  h_{\pi \nu } \partial _{\xi \kappa } h_{\alpha \xi } \partial _{\pi \alpha } h_{\nu \kappa } \\
 - h_{\alpha \xi } \partial _{\nu \pi } h_{\xi \kappa } \partial _{\pi \nu } h_{\kappa \xi }
 - 2h_{\kappa \nu } \partial _{\xi \pi } h_{\alpha \xi } \partial _{\pi \alpha } h_{\nu \kappa }  - h_{\pi \nu } \partial _{\xi \kappa } h_{\alpha \alpha } \partial _{\pi \xi } h_{\nu \kappa } \\
    -  h_{\pi \nu } \partial _{\xi \kappa } h_{\kappa \nu } \partial _{\pi \xi } h_{\alpha \alpha }  + h_{\xi \kappa } \partial _{\nu \pi } h_{\alpha \alpha } \partial _{\pi \nu } h_{\xi \kappa } )
\end{array}
\end{equation}

\begin{equation}
\begin{array}{l}
\delta _1^{(1)} h_{\alpha \beta }  = g(\partial _{\pi \kappa }
\partial _{\kappa \nu } \xi _{\alpha \pi } h_{\nu \beta }  +
\partial _{\kappa \nu } \partial _{\nu \beta } \xi _{\kappa \pi }
h_{\alpha \pi }  - s\partial _{\alpha \pi } \partial _{\kappa \nu
} \xi _{\pi \kappa } h_{\nu \beta } ) + (\alpha  \Leftrightarrow
\beta )
\end{array}
\end{equation}

\begin{equation}
\begin{array}{l}
S_2^{(3)}  = g\int {dx(24\partial _{ o\alpha} \partial _{\nu o}
h_{\alpha\beta} \partial _{\mu\nu} h_{\beta\kappa} \partial
_{\lambda\mu}
 h_{\kappa\lambda}}
+ 12\partial _{ o\alpha} \partial _{\nu o} h_{\alpha\beta} \partial _{\mu\nu} h_{\kappa\lambda} \partial _{\lambda\mu} h_{\beta\kappa}   \\
   - 12\partial _{ o\alpha} \partial _{\nu o} h_{\kappa\lambda} \partial _{\lambda\mu} h_{\alpha\beta} \partial _{\mu\nu} h_{\beta\kappa}  + 24\partial _{\kappa\lambda} \partial _{\mu\nu} h_{\lambda\mu} \partial _{\nu o} h_{\beta\kappa} \partial _{ o\alpha} h_{\alpha\beta}  - 24\partial _{\kappa\lambda} \partial _{\nu o} h_{\alpha\beta} \partial _{ o\alpha} h_{\beta\kappa} \partial _{\mu\nu} h_{\lambda\mu}  \\
   + 24\partial _{\mu\nu} \partial _{\nu o} h_{\alpha\beta} \partial _{ o\alpha} h_{\beta\kappa} \partial _{\kappa\lambda} h_{\lambda\mu}  + 8h_{\alpha\beta} \partial _{ o\alpha} \partial _{\nu o} h_{\mu\nu} \partial _{\kappa\lambda} \partial _{\lambda\mu} h_{\beta\kappa}  + 16\partial _{\kappa\lambda} \partial _{\lambda\mu} h_{\mu\nu} \partial _{\nu o} h_{\beta\kappa} \partial _{ o\alpha} h_{\alpha\beta}   \\
   + 16\partial _{\kappa\lambda} \partial _{\lambda\mu} h_{\mu\nu} \partial _{\nu o} h_{\alpha\beta} \partial _{ o\alpha} h_{\beta\kappa}  + 8\partial _{\kappa\lambda} \partial _{\nu o} h_{\mu\nu} \partial _{\lambda\mu} h_{\beta\kappa} \partial _{ o\alpha} h_{\alpha\beta}  - 8\partial _{\kappa\lambda} \partial _{\nu o} h_{\mu\nu} \partial _{\lambda\mu} h_{\alpha\beta} \partial _{ o\alpha} h_{\beta\kappa}  \\
   + 24\partial _{\kappa\lambda} \partial _{ o\alpha} h_{\beta\kappa} \partial _{\nu o} h_{\alpha\beta} \partial _{\lambda\mu} h_{\mu\nu}  - 12\partial _{\mu\alpha} \partial _{\nu o} h_{\alpha\beta} \partial _{ o\nu} h_{\beta\kappa} \partial _{\lambda\mu} h_{\kappa\lambda}  + 6\partial _{\mu\alpha} \partial _{\nu o} h_{\alpha\beta} \partial _{ o\nu} h_{\kappa\lambda} \partial _{\lambda\mu} h_{\beta\kappa}  \\
   + 6\partial _{\lambda\mu} \partial _{\nu o} h_{\alpha\beta} \partial _{\mu\alpha} h_{\kappa\lambda} \partial _{ o\nu} h_{\beta\kappa}  + 4h_{\beta\kappa} \partial _{ o\nu} \partial _{\nu o} h_{\alpha\beta} \partial _{\mu\alpha} \partial _{\kappa\lambda} h_{\lambda\mu}  - 4\partial _{\mu\alpha} \partial _{\nu o} h_{\alpha\beta} \partial _{ o\nu} h_{\beta\kappa} \partial _{\kappa\lambda} h_{\lambda\mu}   \\
   + 8\partial _{\kappa\lambda} \partial _{\nu o} h_{\alpha\beta} \partial _{\mu\alpha} h_{\lambda\mu} \partial _{ o\nu} h_{\beta\kappa}  + (2 + 2s)h_{\alpha\beta} \partial _{ o\lambda} \partial _{\lambda\mu} h_{\kappa\alpha} \partial _{\mu\nu} \partial _{\nu o} h_{\beta\kappa} \\ + (14 + 2s)\partial _{\lambda\mu} \partial _{\mu\nu} h_{\alpha\beta} \partial _{ o\lambda} h_{\kappa\alpha} \partial _{\nu o} h_{\beta\kappa}  \\
   - 8h_{\alpha\beta} \partial _{ o\kappa} \partial _{\lambda\mu} h_{\alpha\beta} \partial _{\mu\nu} \partial _{\nu o} h_{\kappa\lambda}  + 8\partial _{ o\kappa} \partial _{\nu o} h_{\alpha\beta} \partial _{\lambda\mu} h_{\kappa\lambda} \partial _{\mu\nu} h_{\beta\alpha}  - 8h_{\beta\kappa} \partial _{\lambda\alpha} \partial _{\kappa\lambda} h_{\alpha\beta} \partial _{ o\mu} \partial _{\nu o} h_{\mu\nu}   \\
   + 8\partial _{\kappa\lambda} \partial _{\nu o} h_{\alpha\beta} \partial _{\lambda\alpha} h_{\beta\kappa} \partial _{ o\mu} h_{\mu\nu}  + 32\partial _{\lambda\alpha} \partial _{\nu o} h_{\alpha\beta} \partial _{\kappa\lambda} h_{\beta\kappa} \partial _{ o\mu} h_{\mu\nu}  - 8\partial _{\mu\nu} \partial _{\nu o} h_{\alpha\alpha} \partial _{\lambda\mu} h_{\beta\kappa} \partial _{ o\lambda} h_{\beta\kappa}  \\
   + 4\partial _{\lambda\mu} \partial _{\nu o} h_{\alpha\alpha} \partial _{\mu\nu} h_{\beta\kappa} \partial _{ o\lambda} h_{\beta\kappa}  - (1 + s)h_{\kappa\alpha} \partial _{\lambda\mu} \partial _{\mu\lambda} h_{\alpha\beta} \partial _{ o\nu} \partial _{\nu o} h_{\beta\kappa} \\ - (\frac{5}
{2} + s)\partial _{\lambda\mu} \partial _{\mu\lambda} h_{\alpha\beta} \partial _{\nu o} h_{\beta\kappa} \partial _{ o\nu} h_{\kappa\alpha}   \\
   + 4h_{\alpha\beta} \partial _{\nu o} \partial _{ o\nu} h_{\beta\alpha} \partial _{\mu\kappa} \partial _{\lambda\mu} h_{\kappa\lambda}  + 6\partial _{\lambda\mu} \partial _{\mu\kappa} h_{\kappa\lambda} \partial _{\nu o} h_{\alpha\beta} \partial _{ o\nu} h_{\beta\alpha}  + 8\partial _{\mu\kappa} \partial _{\nu o} h_{\alpha\beta} \partial _{\lambda\mu} h_{\beta\alpha} \partial _{ o\nu} h_{\kappa\lambda}  \\
   + 2\partial _{ o\nu} \partial _{\nu o} h_{\kappa\kappa} \partial _{\lambda\mu} h_{\alpha\beta} \partial _{\mu\lambda} h_{\beta\alpha})  \\
\end{array}
\end{equation}

\begin{equation}
\begin{array}{l}
\delta _2^{(1)} h_{\alpha \beta }  = g(\xi _{\alpha \kappa }
h_{\kappa \beta }  + \xi _{\beta \kappa } h_{\kappa \alpha } )
\end{array}
\end{equation}

where $g$ is an interaction constant. There are many possible
forms of rewriting this solution, by using identities between
different third-order terms in second action, presented form seems
to be most compact.

\section{Conclusion}
Having symmetry variations in non-zero over field order, one
usually obtains a non-Abelian  algebra of symmetries. However, in
our case we have an obstacle in calculating an algebra of gauge
transformations. In usual case symmetry statement (\ref{eq3})
include one term, and commutator of symmetries is again a
symmetry. When number of terms is more than one, as is in our
case, it is easy to understand, that commutator of symmetries is
no longer a symmetry. We are not aware on a procedure (which
should exist, nevertheless) of obtaining third symmetry from two
existing symmetries of type (\ref{eq3}) with more than one action.
It is worth to mention that one other property, namely, the fact
that symmetry of equations of motion follows from that of action,
is maintained in this generalized case. That can be simply proved
by differentiating (\ref{eq3}).

There are few directions of development of the present results.
Next order terms can be derived, in principle, in the same way,
but will require too much calculations, so we need an
understanding of symmetry formulae, e.g. whether some geometry
exists behind these symmetries.  One can try to construct the
interaction terms for few spin 1 fields, i.e. generalize the
Yang-Mills theory. These calculations also are sufficiently
complicated. The quantization of these models is another problem.
Approach developed in \cite{Vas}, particularly the notion of
positive and negative modes, seems to be applicable in above case,
also. The existence of actions in the present approach perhaps
should provide a possibility of generalization of path integral
quantization.

\section{Acknowledgements}

This work is supported partially by INTAS grant \#99-1-590. I'm
indebted to R.Manvelyan for discussions.


\begin{thebibliography}{9}

\bibitem{Tow}
 P.K. Townsend, $M$-theory from its superalgebra,
hep-th/9712004.
\bibitem{Man1}  R. Manvelyan and R. Mkrtchyan, Towards SO(2,10)-invariant
M-theory: Multilagrangian Fields,  Mod.Phys.Lett. A15 (2000)
747-760, hep-th/9907011.
\bibitem{Bars} I. Bars, Supersymmetry, p-brane duality and hidden space
and time dimensions, hep-th/9604139, Phys.Rev. D54 (1996)
5203-5210 \newline I. Bars, Duality and hidden dimensions,
hep-th/9604200,\newline I.Bars, S-Theory, hep-th/9607112 Phys.Rev.
D55 (1997) 2373-2381.

\bibitem{Man2}  R. Manvelyan and R. Mkrtchyan, Free Field Equations For Space-Time Algebras With Tensorial
Momentum, Mod. Phys. Lett. A, Vol. 17, No. 21 (2002) pp.1393-1406,
hep-th/0112233.

\bibitem{Rud} I. Rudychev and E. Sezgin, Phys. Lett. \textbf{B415} (1997) 363
\bibitem{Band} I.Bandos  and J.Lukierski, Tensorial Central Charges and New Superparticle
Models with Fundamental Spinor Coordinates, hep-th/9811022,
Mod.Phys.Lett. A14 (1999) 1257-1272 \newline I.Bandos, J.Lukierski
and D.Sorokin, Superparticle Models with Tensorial Charges, Phys.
Rev. \textbf{D61} (2000) 45002, hep-th/9904109.


\bibitem{Vas} M.A.Vasiliev, Relativity, Causality, Locality,
Quantization and Duality in the Sp(2M) Invariant Generalized
Space-Time, hep-th/0111119, Conformal Higher Spin Symmetries of 4d
Massless Supermultiplets and osp (L,2M) Invariant Equations in
Generalized (Super)Space, hep-th/0106149.
\bibitem{Mkr} R.Mkrtchyan, unpublished.

\end{thebibliography}
\end{document}